# The thermodynamic equilibrium of gas in a box divided by a piston

**Alberto Herrera-Gomez**
*CINVESTAV-Querétaro. Querétaro, México 76230.*

The equilibrium conditions of a system consisting of a box with gas divided by a piston are revised. The apparent indetermination of the problem is solved by explicitly imposing the constancy of the internal energy when the Entropy Maximum Principle is applied. The equality of the pressures is naturally concluded from this principle when the piston is allowed to spontaneously move. The application of the Energy Minimum Principle is also revised.

## I. Introduction

The extraction from the thermodynamic principles of the equilibrium conditions for a system consisting of a gas-containing box divided by an adiabatic piston has been an elusive problem.[1,2,3] It is clear that the mechanical equilibrium is reached when the pressures in both sides are equal to each other. However, how to extract this condition from the Entropy Maximum Principle has proven to be confusing. Some authors have claimed that the problem is indeterminate.[1,4] Some others have even got to the (wrong) conclusion that the equilibrium is reached when the ratio of the pressure to the temperature $P/T$ is the parameter that should be equal in both sides.[5,6] Many other authors also arrived to this same conclusion but adding the statement that the temperatures should also be equal to each other, thus effectively considering only the case of a diathermal wall.[4,7,8]

In 1969 Curzon treated this problem and tried to conclude the equality of the pressures from the maximization of the entropy.[9] This work is frequently referred as the appropriate approach.[1,3,10] However, the argument employed assumes that the increase of entropy is identically equal to zero, which, as discussed below and pointed out earlier,[11] is equivalent to assume (not show) that the pressures are equal. It also assumes an incorrect relationship between the change in volume and the change in internal energy.

All the approaches employ the following relation. It can be directly shown that the change of total entropy $S$ when energy, volume, and particles (of type $a$) can be interchanged between the two sides of the box is given by

$$dS = \left(\frac{1}{T_1} - \frac{1}{T_2}\right)dU_1 + \left(\frac{P_1}{T_1} - \frac{P_2}{T_2}\right)dV_1 - \left(\frac{\mu_{a,1}}{T_1} - \frac{\mu_{a,2}}{T_2}\right)dN_{a,1},$$

where $U_i$, $T_i$, $P_i$, $V_i$, $\mu_{a,i}$, and $N_{a,i}$ ($i=1,2$) are the internal energy, temperature, pressure, volume, chemical potential of specie $a$, and number of particles of specie $a$ for each side. The system is assumed isolated ($dU_1 + dU_2 = 0$), with constant total volume ($dV_1 + dV_2 = 0$), and with constant number of particles ($dN_{a,1} + dN_{a,2} = 0$). From this equation it is indeed possible to extract the condition of thermal equilibrium. By removing the constraint of heat flow (diathermal piston), $dU_1$ could be made different from zero and at the same time keeping the volume and number of particles fixed in each side ($dV_1 = 0$ and $dN_{a,1} = 0$). Since in equilibrium $S$ is maximum ($dS = 0$), then the temperatures should be equal when the piston wall is diathermal.

Part of the existing confusion arises when the condition of mechanical equilibrium is tried to be extracted by making zero the coefficient of $dV_1$. It should be stressed that a change on $V_1$ implies a change on $U_1$. Since they are not independent, the whole combination $(1/T_1 - 1/T_2)dU_1 + (P_1/T_1 - P_2/T_2)dV_1$ should instead be set to zero. It is indeed possible to extract the equality of the pressures from this condition. This is properly done in the following section.

In section III and V it is discussed the wording of the Entropy Maximum and Energy Minimum Principles. The extraction of the equilibrium conditions by employing the Energy Minimum Principle is discussed in Section VI. The application of the proposed approach to the solution of some simple problems is provided in Section IV.

## II. The equilibrium conditions obtained from the Entropy Maximum Principle

Another part of the confusion is the wide-spread error of applying the Entropy Maximum Principle without assuring that the total internal energy is kept constant. It is often assumed that the change on internal energy in each side is $-P_i\,dV_i$ ($i = 1,2$).[9] If this were the case, the total energy would not be constant since the change on the total energy due to a displacement of the piston would be $-(P_1 - P_2)dV_1$. The Entropy Maximum Principle is only valid when the total energy is constant.[12,4] In order to keep the total energy constant it is necessary to give back this energy to the system. There are various ways to do this. One possibility is to accelerate the piston and provide it with kinetic energy. However, if there is any friction, as small as it might be, the kinetic energy will eventually be converted into heat. Another possibility is to consider a process in which this work is quasistatically given back to the system. This could be achieved by allowing the friction to do its job while the piston moves, or by waiting until the vibrational energy generated by the piston striking a new constrain turns into heat. In any case, the work is converted in an amount of heat equal to

$$dQ = (P_1 - P_2)dV_1.$$

This assures that the system is kept in a surface of constant energy. It should be noted that the entropy increases precisely because the available energy (work) is converted into heat. In fact, the Entropy Maximum Principle of thermodynamics is sometimes stated as[13]

> *If a closed system is at some instant in a non-equilibrium macroscopic state, the most probable consequence at later instants is a steady increase in the entropy of the system.*

and[6]

> *The spontaneous tendency of a system to go towards equilibrium cannot be reversed without at the same time changing some organized energy, work, into disorganized energy, heat.*

How much of this heat ends in side 1 or in side 2 depends on the details of the quasistatic process and of the system. In general, a fraction $\chi$ ends in side 1 and the rest $(1-\chi)$ in side 2; then

**Eq. 1**  $dQ_1 = \chi\, dQ + dQ_{2\to 1}$  and  $dQ_2 = (1-\chi)\, dQ + dQ_{1\to 2}$

where $dQ_i$ ($i=1,2$) is the heat incorporated into side *i*, $dQ_{2\to 1}$ is the heat transferred into side 1 from side 2 (through the piston wall), and $dQ_{1\to 2}$ is the opposite. By definition, $dQ_{2\to 1} + dQ_{1\to 2} = 0$. The total change of entropy is then

$$dS = dS_1 + dS_2 = \frac{\chi\, dQ + dQ_{2\to 1}}{T_1} + \frac{(1-\chi)\, dQ + dQ_{1\to 2}}{T_2}.$$

By employing Eq. 1 it is possible to write this equation in the following way:

**Eq. 2**  $dS = \left(\dfrac{1}{T_1} - \dfrac{1}{T_2}\right) dQ_{2\to 1} + \left(\dfrac{\chi}{T_1} + \dfrac{1-\chi}{T_2}\right)(P_1 - P_2)\, dV_1.$

Note that the factor containing $\chi$ is never zero or negative for $0 \le \chi \le 1$. Since the total internal energy has been kept constant, it is possible to apply the Entropy Maximum Principle. With Eq. 2 it is now possible to establish the conditions of equilibrium. It should be stressed that the amount of heat transferred between the sides ($dQ_{2\to 1}$) and the change of volume ($dV_1$) are independent from each other. They area controlled by independent constraints.

It is now possible to extract the thermodynamic equilibrium conditions. Since at equilibrium $dS = 0$, then either the temperatures are equal or the heat transfer is not allowed ($dQ_{2\to 1} = 0$). In addition, either the pressures are equal or the exchange of volume is not allowed ($dV_1 = 0$). As a conclusion,

> *the equality of the pressures when the piston is allowed to move is an equilibrium condition naturally obtained from the of the Entropy Maximum Principle.*

In the same way, it can be shown that, when the exchange of particles is allowed, the change on entropy is given by

**Eq. 3**  $dS = \left(\dfrac{1}{T_1} - \dfrac{1}{T_2}\right) dQ_{2\to 1} + \left(\dfrac{\chi}{T_1} + \dfrac{1-\chi}{T_2}\right)(P_1 - P_2)\, dV_1 - \left(\dfrac{\chi}{T_1} + \dfrac{1-\chi}{T_2}\right)(\mu_{a,1} - \mu_{a,2})\, dN_{a,1},$

Then, it can be stated that

> *the equality of the chemical potentials (when particle exchange is allowed) is also a consequence of the Entropy Maximum Principle.*

It is important to emphasize this because it has wrongly been claimed that the chemical equilibrium (when the flow of particle of type *a* is allowed) is achieved when the ratio of the chemical potential to the temperature ($\mu_a / T$) is the same in both sides.[6]

## III. A note about the statement of the Entropy Maximum Principle

The precise way in which the Entropy Maximum Principle is enunciated can contribute to the confusion described in Section I. Some authors make special emphasis on internal parameters. Examples are the statements employed by Callen[4] *the values assumed by the extensive parameters in the absence of an internal constraint are those that maximize the entropy over the manifold of constrained equilibrium states,*

and

*the equilibrium value of any unconstrained internal parameter is such as to maximize the entropy for the given value of the total internal energy.*

It is clear that $V_1$ and $N_{a1}$ qualify as extensive parameters. However, the quantity $Q_{2\to 1}$ associated to the other differential $dQ_{2\to 1}$ in Eq. 3 does not. Since $dQ_{2\to 1}$ represents a variation, there is no discrepancy if the Gibbs text is employed instead:[12]

*For the equilibrium of any isolated system it is necessary and sufficient that in all possible variations of the state of the system which do not alter its energy, the variation of its entropy shall either vanish or be negative.*

A more operative statement which is equivalent to the previous one is the following:

***The spontaneous changes in an isolated system when a constraint is removed are such that the entropy increases, reaching a maximum value at equilibrium.***

The changes of the system are represented by $dQ_{2\to 1}$, $dV_1$, and $dN_{a,1}$. Away from equilibrium, their signs are such that the total entropy increases. For example, if $T_2 > T_1$ the sign of $dQ_{2\to 1}$ should be positive to cause an increase on the total entropy. The exchange of heat, volume, and particles are the result of the removal of independent constrains. Conceptually, it is clear how each one could be independently removed.

To illustrate how the use of Callen's statement of the Entropy Maximum Principle could lead to wrong conclusions, the Eq. 2 will be written in terms of differentials of internal variables. Since

$$dS_1 = \frac{\chi\, dQ + dQ_{2\to 1}}{T_1} = \frac{\chi(P_1 - P_2)dV_1 + dQ_{2\to 1}}{T_1},$$

then

$$dQ_{2\to 1} = T_1\, dS_1 + \chi(P_1 - P_2)dV_1.$$

Equation .2 could then be written as

**Eq. 4** $\quad dS = \left(1 - \dfrac{T_1}{T_2}\right)dS_1 + \dfrac{(P_1 - P_2)}{T_2}\, dV_1.$

If particle exchange is allowed, then

**Eq. 5** $\quad dS = \left(1 - \dfrac{T_1}{T_2}\right)dS_1 + \dfrac{(P_1 - P_2)}{T_2}\, dV_1 - \dfrac{(\mu_{a,1} - \mu_{a,2})}{T_2}\, dN_{a,1}.$

(Note that $dS$ is not longer explicitly dependent on $\chi$; its dependence is implicit through $dS_1$.) In this way the differential on total entropy is written in term of the differentials of the internal parameters $S_1$, $V_1$, and $N_{a,1}$. Although it is tempting to try to extract the equilibrium conditions from Eq. 4 (or Eq. 5), this cannot be done because the differentials $dS_1$ and $dV_1$ (and $dN_{a,1}$) are not independent from each other. Experimentally, they cannot be turned on and off independently by removing the appropriate constraints. As shown above, the adiabatic condition of the piston wall does not imply that $dS_1 = 0$. It is very important to note that Eq. 4 does not imply that the change on total entropy caused by a change on $V_1$ is equal to $(P_1 - P_2)/T_2 \, dV_1$ since a change in $V_1$ causes a change on $S_1$ ($dS_1$ is in general different from zero, it depends on $\chi$, which is not controlled by the constrains).

This line of thought could be taken further down. If $dS_1$ is written in terms of $dU_1$, then it is possible to return to the expression

$$dS = \left(\frac{1}{T_1} - \frac{1}{T_2}\right) dU_1 + \left(\frac{P_1}{T_1} - \frac{P_2}{T_2}\right) dV_1 - \left(\frac{\mu_{a,1}}{T_1} - \frac{\mu_{a,2}}{T_2}\right) dN_{a,1}.$$

Since the differentials $dU_1$, $dV_1$, and $dN_{a,1}$ cannot be turned on and off independently from each other by removing the appropriate constraints, the equilibrium principle under the removal of constrains does not imply that the coefficients could be independently set to zero. In fact, doing so would yield to wrong conclusions (such as that the mechanical equilibrium is reached when $P_1/T_1 = P_2/T_2$). However, the terms could be rearranged as in Eq. 3, where the differentials are actually independent from each other, and the correct equilibrium conditions could be found from the Entropy Maximum Principle.

## IV. The equilibrium conditions obtained from the Energy Minimum Principle

In Callen's classic book of thermodynamics, the conditions of equilibrium of the system here in question are attempted to be extracted from the Energy Minimum Principle.[4] However, it is not done properly. The error consisted of not assuring that the entropy remains constant. This is fundamental since the principle of minimum energy applies only when the total entropy is kept constant.[4,6,12] The quasistatic exchange of heat through the wall induces the following change on total entropy

**Eq. 6** $\quad dS_{\text{exchange}} = \left(\frac{1}{T_1} - \frac{1}{T_2}\right) dQ_{2 \to 1}.$

Since heat flows from the warm side to the cooler one, $dS_{\text{exchange}}$ is greater than zero if the temperatures are different. Requiring that $dS_{\text{exchange}} = 0$ when heat exchange is allowed is equivalent to require that the temperatures are the same. In this way, the equality of the temperatures is assumed, not shown.

It is possible to show that the temperatures should be equal in equilibrium in the following way. To actually keep the entropy constant (when the temperatures are not equal and heat transfer is allowed) it is necessary to extract certain amount of heat $dQ$ from the system, a fraction $\chi$ from side 1 and the rest from side 2. Then, the total change in entropy is

$$dS = \left(\frac{1}{T_1} - \frac{1}{T_2}\right)dQ_{2\to 1} + \frac{\chi}{T_1}dQ + \frac{1-\chi}{T_2}dQ.$$

If the entropy is to be kept constant, the amount of heat that should be extracted is

$$dQ = \frac{1}{\frac{\chi}{T_1} + \frac{1-\chi}{T_2}}\left(\frac{1}{T_2} - \frac{1}{T_1}\right)dQ_{2\to 1}.$$

The total change on internal energy is then

**Eq. 7** $$dU = \frac{1}{\frac{\chi}{T_1} + \frac{1-\chi}{T_2}}\left(\frac{1}{T_2} - \frac{1}{T_1}\right)dQ_{2\to 1} + (P_1 - P_2)dV_1 + (\mu_{a,1} - \mu_{a,2})dN_{a,1}.$$

This assures that the system evolves in a surface of constant entropy. The principle of minimum energy can now be applied. It is then direct to show that, in equilibrium, if heat exchange is allowed, the temperatures should be the same, and that, if volume is allowed to change, the pressures should be the same. In addition, if particle exchange is allowed, then in equilibrium the chemical potentials are equal.

The equality of the temperatures, pressures, and chemical potentials when heat can flow, volume can change, and particles can permeate, respectively, is reached regardless the total internal energy is kept constant (employing the Entropy Maximum Principle) or the total entropy is kept constant (employing the Energy Minimum Principle).

## V. A note about the statement of the Energy Minimum Principle

The way in which Callen enunciates the Energy Minimum Principle:[4]

*the equilibrium value of any unconstrained internal parameter is such as to minimize the energy for the given value of the total entropy,*

also emphasizes on the minimization respect to internal parameters. This statement is not operative since $Q_{2\to 1}$ is not an internal parameter, so it cannot be employed to extract the equilibrium conditions from Eq. 7. There is no problem if Gibbs statement is employed:[12]

*For the equilibrium of any isolated system it is necessary and sufficient that in all possible variations of the state of the system which do not alter its entropy, the variation of its energy shall either vanish or be positive*

A more functional way to enunciate the Energy Minimum Principle which is consistent with the previous one is the following:

***The spontaneous changes not causing a variation on the total entropy of a system when a constraint is removed are such that the internal energy decreases, reaching a minimum value at equilibrium.***

## VI. Solution to some special cases

The approach used in Sections II and IV is operative in the sense that it paves the way for the prediction of the final state, which is the main objective of thermodynamics. Some simple cases will be described. In these examples the gas will be assumed as ideal and monatomic, the process as quasistatic, and the wall of the piston as adiabatic. For each side it holds that

**Eq. 8**    $p = \dfrac{2}{3}\dfrac{U}{V}$  and  $T = \dfrac{2}{3}\dfrac{U}{RN}$ .

### A.   Perfect adiabatic subsystems

The simplest case is in which the work $-(P_1 - P_2)dV_1$ is not given back to the system in the form of heat or in any form. This problem could be solved in a different way, but it is illustrative for describing the procedure. In this case the entropy of each side is kept constant, thus $dU_i = -P_i\,dV_i\ (i=1,2)$. This equation could be solved to show that $P_i V_i^{5/3}\ (i=1,2)$ remains constant for each side (a known relation for adiabatic systems). Since that total entropy remains constant, it is possible to apply the principle of minimum energy and require that $P_{1,f} = P_{2,f}$. It is straightforward to show that the final volume can be calculated from the initial volumes and pressures as follows

$$V_{1,f} = \dfrac{P_{1,0}^{3/5} V_{1,0}}{P_{1,0}^{3/5} V_{1,0} + P_{2,0}^{3/5} V_{2,0}}\, V_{tot}\,.$$

The final pressures, temperatures, and energies could then be calculated employing Eq. 8.

### B.   Slow piston (no oscillation)

When the friction is large enough to prevent the acceleration of the piston, the work $-(P_1 - P_2)dV_1$ is completely transformed into heat, warming up the enclosing cylinder and the border of the piston in contact with the cylinder. The gas in the expanding side becomes in contact with the region of the cylinder wall that has just been swept by the piston border. If the transfer of this heat to the gas is faster than the diffusion of that heat within the cylinder wall (and from the piston border to the rest of the piston), then most of the heat will be gained by the expanding side. When $P_1 > P_2$, this could be represented by a value of $\chi$ close to 1. The system evolves according to the following equations

$$dU_1 = \chi(P_1 - P_2)dV_1 - P_1\,dV_1$$
$$dU_2 = (1-\chi)(P_1 - P_2)dV_1 + P_2\,dV_1\,.$$

Since the total energy is kept constant, it is possible to use the maximum entropy principle and require that the final pressures are equal to each other. This system of equations could be solved analytically if $\chi$ is assumed to be constant. The extreme case is that in which all the heat is transferred into one side, which is equivalent to assume that $\chi = 1$. For this particular case it is straightforward to show that the final volume of side two is given by

$$V_{2,\mathrm{f}} = \left(\frac{U_{2,0} V_{\mathrm{tot}}}{U_{\mathrm{tot}}}\right)^{2/5} V_{2,0}^{3/5}.$$

The final pressures, temperatures, and energies could then be calculated employing Eq. 8.

### C. Oscillating piston

If the piston accelerates appreciably, its kinetic energy should be part of the balance. The equation of motion of the piston is

$$\sigma A \frac{d^2 v}{dt^2} = (P_1 - P_2)A - \kappa v.$$

where $\sigma$ is the surface density of the piston, $A$ is the area, $\kappa$ is the friction coefficient, and $v$ is the speed of the piston. The energy converted into heat as the piston moves is $\kappa v \, dx$, which is always positive. Since the acceleration of the piston is not zero, this energy is not longer equal to $(P_1 - P_2)dV_1$. The internal energy of each side evolves in the following way

$$dU_1 = \begin{cases} \chi \kappa v \, dx - P_1 \, dV_1 & \text{if } 0 \leq v \\ (1-\chi)\kappa v \, dx - P_1 \, dV_1 & \text{if } v < 0 \end{cases}$$

$$dU_2 = \begin{cases} (1-\chi)\kappa v \, dx + P_2 \, dV_1 & \text{if } 0 \leq v \\ \chi \kappa v \, dx + P_2 \, dV_1 & \text{if } v < 0 \end{cases}$$

where $dx$ is the piston displacement ($dV_1 = A \, dx$). This set of three equations could only be solved numerically even if $\chi$ is assumed to be constant.

A more precise functional dependence of $\chi$ could be calculated from kinetic theory for a particular system. It is also possible to treat $\chi$ as a parameter for fitting experimental data.

## VII. Conclusions

The Entropy Maximum Principle was applied to a system consisting of a gas containing box divided by a piston. The determination of the equilibrium conditions employing this principle has been a long open problem of thermodynamics. The discrepancies arose from a generalized error consisting of not assuring that the energy remains constant when the piston moves. The energy can only stay constant if the energy lost in the form of work is returned in the system in the form of heat. This is the natural way in which the entropy increases. When the energy is actually kept constant, the equilibrium conditions follow naturally. An important conclusion is that the equality of the pressures of both sides when the piston is allowed to move is an equilibrium condition directly obtained from the Entropy Maximum Principle. The equality of the chemical potentials when particles could be interchanged is also a natural consequence of the Entropy Maximum Principle.

Part of the heat returned to the system is incorporated into one of the two sides and the rest into the other side. This is quantified by the introduction of the parameter $\chi$. The proposed approach provides a means to solve particular problems in terms of this parameter. For cases in which $\chi$ could be considered constant, it is possible to find analytical solutions. For numerical analysis of experimental data $\chi$ could be treated as a fitting parameter. More precise determinations of $\chi$ could involve the application of kinetic theory.

Although not recognized, another long standing problem of thermodynamics is the extraction of the equilibrium conditions form the Energy Minimum Principle. A common error is trying to apply the principle without assuring that the entropy remains constant. If this is incorporated into the calculations, it is possible to show, without circular arguments or without first assuming it, that the temperatures should be equal in both sides if heat flow is allowed. The equality of the temperatures, pressures, and chemical potentials when heat can flow, volume can change, and particles can permeate, respectively, is reached regardless the total internal energy is kept constant (employing the Entropy Maximum Principle) or the total entropy is kept constant (employing the Energy Minimum Principle).

The statements of the Entropy Maximum Principle and of the Energy Minimum Principle should not be done in terms of the internal parameters but in term of internal variations or changes.

## Acknowledgements

The author acknowledges the students of the course of thermodynamics (2010, at Cinvestav-Queretaro) for bearing with him during the development of the approach proposed in this paper.